1101# Navigation of a UAV Equipped with a Reconfigurable Intelligent Surface for LoS Wireless Communication with a Ground Vehicle

Mohsen Eskandari, Hailong Huang, Andrey V. Savkin, Wei Ni

*Abstract*— Unmanned aerial vehicles (UAVs) have been successfully adopted to enhance the flexibility and robustness of wireless communication networks. And recently, the reconfigurable intelligent surface (RIS) technology has been paid increasing attention to improve the throughput of the fifth-generation (5G) millimeter-wave (mmWave) wireless communication. In this work, we propose an RIS-outfitted UAV (RISoUAV) to secure an uninterrupted line-of-sight (LoS) link with a ground moving target (MT). The MT can be an emergency ambulance and need a secure wireless communication link for continuous monitoring and diagnosing the health condition of a patient, which is vital for delivering critical patient care. In this light, real-time communication is required for sending various clinical multimedia data including videos, medical images, and vital signs. This significant target is achievable thanks to the 5G wireless communication assisted with RISoUAV. A two-stage optimization method is proposed to optimize the RISoUAV trajectory limited to UAV motion and LoS constraints. At the first stage, the optimal tube path of the RISoUAV is determined by taking into account the energy consumption, instant LoS link, and UAV speed/acceleration constraints. At the second stage, an accurate RISoUAV trajectory is obtained by considering the communication channel performance and passive beamforming. Simulation results show the accuracy and effectiveness of the method.

## I. Introduction

THE human societies have been shifting toward modernity owing to the fast development of information and communication technology and Internet-of-things (IoT). The realization of smart cities is probably the vision of this trend which is designed based on the high-performance wireless communication infrastructure [1]. In this sense, the International Telecommunication Union (ITU) has projected 5 zettabytes (ZB)/month global mobile data transfer [2]. This has motivated research efforts to hop from fifth-generation (5G) mobile communication to the sixth generation (6G) broadband cellular networks [4]. However, the overall loss of the millimeter-wave (mmWave) communication, which is the crucial technology of 5G, is larger than that of a microwave system for a point-to-point link regarding the signal propagation due to higher path loss and fading, refraction/diffraction, blockage, etc. [4]. Thus, efforts have been focused on solving problems with mmWave communication with massive multi-input multi-output (MIMO) base stations (BSs), to secure ubiquitous connectivity.

Regarding the small wavelength of electromagnetic waves in mmWave systems ($\lambda = 1{\sim}10$ mm), the reconfigurable intelligent surface (RIS) technology has been developed to enhance the flexibility and robustness of wireless communication networks exposed to coverage holes, blind spots, and obstacles [5]. RISs comprise sub-wavelength meta-material unit cells that can be controlled/programmed to alter the amplitude and phase of the incident electromagnetic mmWave signal (in a real-time reconfigurable manner) [6]. Thanks to their intelligent reconfigurability, RISs can produce favorable channel conditions for mmWave wireless communication by absorbing and then reflecting (beamforming or beam steering) the transmitted signal at the receiver side [7]-[8].

In the meanwhile, unmanned aerial vehicles (UAVs) have gained significant attention to improve the overall spectrum efficiency of wireless communications owing to their great 3D mobility [9]. The emerging RIS-outfitted UAV (RISoUAV) technology (UAV with RIS) has been getting increasing attention for further enhancing the flexibility of the new generations of wireless communication networks by establishing clear/instant line-of-sight (LoS) links [5], [10].

In the UAV-assisted wireless communication systems, a UAV acts as an agile aerial relay [11] to provide an indirect link between the BS and user equipment and improve the achievable rate [12]. For instance, the RISoUAV is adopted as a passive aerial relay for sampled information by IoT devices [13]. The UAV position, alignment, and trajectory can be optimized to improve system efficiency and performance and to preserve energy [14]. Due to the limited onboard energy at UAVs, the consumed energy by the UAVs should be preserved. To this

This work was supported by the Australian Research Council. Also, this work received funding from the Australian Government, via grant AUSMURIB000001 associated with ONR MURI grant N00014-19-1-2571.

M. Eskandari and A. V. Savkin are with The School of Electrical Engineering and Telecommunications, University of New South Wales, Sydney, NSW 2052, Australia (e-mail: m.eskandari@unsw.edu.au; a.savkin@unsw.edu.au).

H. Huang is with the Department of Aeronautical and Aviation Engineering, the Hong Kong Polytechnic University, Hong Kong. (E-mail: hailong.huang@polyu.edu.hk).

W. Ni is with Data61, CSIRO, Australia. (E-mail: wei.ni@data61.csiro.au).

end, the minimum energy consumption should be attained by which the maximum spectrum quality (by instant uninterrupted LoS) is achieved [15]. Channel estimation/performance and RIS elements control is another research topic [9].

RIS-assisted secure vehicular wireless communication is also an emerging application that has been considered by optimal placement of the RIS along the highway for vehicle-to-everything (V2X) communication [16]. From another perspective, the secrecy outage performance of RIS-aided vehicle-to-vehicle (V2V) communications has been analyzed in [17].

To the best of the authors' knowledge, this is the first paper that considers the application of an RISoUAV to establish a BS-RIS-mobile vehicle link. The vehicle is an ambulance carrying a patient whose condition must be continually monitored to diagnose the patient remotely (to speed up treatment and increase rescue rate) which needs a real-time connection to emergency management computer systems. Besides, the health conditions of the patient can become critical if they are not evaluated and treated in time. Thus, reliable and real-time communication is critical for safe and effective care during the emergency patient transfer from rural areas to hospitals. The real-time monitoring of patients involves the transmission of various clinical multimedia data including videos, medical images, and vital signs, which requires the use of the mobile network with high-fidelity communication bandwidth [18]. With the aid of the 5G broadband cellular network and RIS technology, it is possible to ensure critical emergency service at a moving smart ambulance [19]; see Fig. 1. This target is achievable since the reliability of the RIS-communication has been analyzed [5] and validated through experimental tests by successfully modulating/transmitting a digital video file through an RIS-based transmitter [20].

The optimal trajectory of the RISoUAV is designed taking into account the energy efficiency, UAV maneuver constraints, LoS link, and communication channel performance. This is a multilateral problem regarding the UAV navigation and its flight requirement/limits, energy consumption of the UAV, nonconvex optimization problem and its convergence to a valid solution, computational burden, RIS beamforming, and channel performance, etc. In this light, we propose a two-stage heuristic optimization scheme by utilizing the rapidly exploring random tree (RRT) algorithm and model predictive control (MPC). At the first stage, the energy consumption, instant LoS link, and navigation issues are considered. The optimal energy-efficient tube path is planned at this stage, based on the map route of the mobile target (MT), i.e., the ambulance, maneuver limits associated with the UAV (speed/acceleration constraints), and the environment 3D plan (with potential blockage/obstacles) to secure LoS. The tube path provides permitted space for the UAV flight through which the maneuver constraints of the UAV are satisfied. Then, at the second stage, an accurate UAV trajectory is obtained through the tube path considering the communication channel performance of the BS-RIS-MT link. To obtain the instant/optimal UAV position, maximizing the achievable rate by the passive beamforming is considered through channel estimation and RIS reconfiguration.

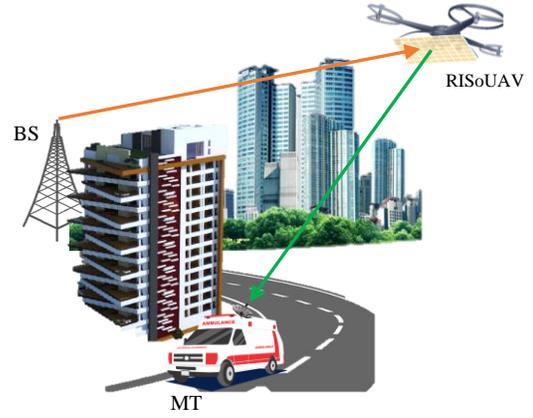

Fig. 1. The mobile RIS-outfitted UAV to hold instant LoS communication with a mobile target (MT) (vehicle).

The remainder of this paper is organized as follows. In Section II, the problem statement is presented along with the modeling and problem formulation. In Section III, the proposed method for optimizing the 3D trajectory of the RISoUAV is presented. Computer simulations are conducted in Section IV to evaluate the performance of the proposed method. Finally, Section V concludes the paper.

## II. PROBLEM STATEMENT, MODELING, AND FORMULATION

The scenario under study is a flying RISoUAV (that carries RISs and flies in the 3D space) acts as a passive aerial relay to ensure that the ground MT is continuously connected to the BSs of the wireless network (where the direct BS-MT LoS link might be blocked by buildings, hills, etc.).

### A. UAV Motion and Energy Consumption Model

UAV agility is of concern in this application. Thus, the RISoUAV motion is modeled by the following nonlinear dynamic model:

$$\begin{gathered} p(\tau) \coloneqq [x(\tau), y(\tau), z(\tau)], \\ \begin{cases} \dot{x}(t) = v(t)\cos(\theta(t)), \\ \dot{y}(t) = v(t)\sin(\theta(t)), \\ \dot{z}(t) = u(t), \\ \dot{\theta}(t) = \omega(t), \end{cases} \\ 0 \le v(t) \le V_{max} \\ -W_{max} \le \omega(t) \le W_{max} \\ -U_{max} \le u(t) \le U_{max} \\ Z_{min} \le z(t) \le Z_{max} \end{gathered} \quad (1)$$

where $p(\tau) \in \mathbb{R}^3$ denotes the UAV position in the Cartesian coordinates at time $\tau$; $v(t)$, $u(t)$, and $\omega(t)$ are linear horizontal, vertical and angular speeds, respectively, that are limited to $V_{max}$, $U_{max}$, and $W_{max}$ that are positive constants indicating the maximum linear horizontal, vertical and angular speeds, respectively; $\theta(t)$ is the heading (yaw angle) of the RISoUAV with respect to the $x-$axis; $Z_{min}$ and $Z_{max}$ denote the minimum and maximum limits of the RISoUAV altitude, respectively. The roll and pitch (elevation) angles are assumed to be fixed due to the maneuver limitation exposed by the RIS.

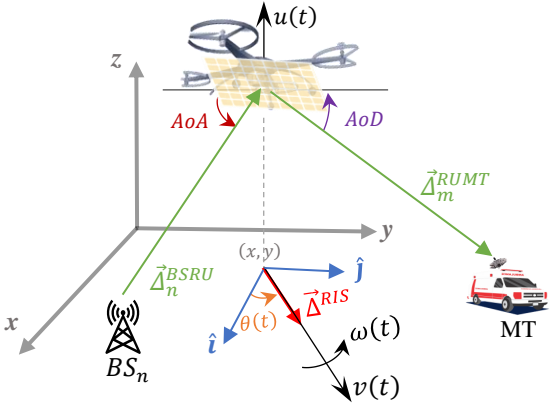

Fig. 2. The RISoUAV position and motion in the Cartesian coordinates.

Here $U(t) = [v_x(t), v_y(t), u(t), \omega(t)]$ is considered as the input to the system in (1), where $v_x(t) = v(t)\cos(\theta(t))$ and $v_y(t) = v(t)\sin(\theta(t))$. Thus, the following energy cost function can be considered for the RISoUAV navigation:

$$E_{RISoUAV} = \int_{t_0}^{t_0+T} \Big(\alpha_1\big(v_x(t) + v_y(t) + |u(t)| + |\omega(t)|\big) \\ + \alpha_2\big(|\dot{v}_x(t)| + |\dot{v}_y(t)| + |\dot{u}(t)| \\ + |\dot{\omega}(t)|\big)\Big) dt \quad (2)$$

where $E_{RISoUAV}$ denotes the energy consumption of the UAV, $T$ denotes the time interval of interest in the $[t_0, \ t_0 + T]$ interval, and $\alpha_1 > 0$ and $\alpha_2 > 0$ are weighting coefficients that compromise the control inputs and UAV maneuver (hovering and propulsion).

*B. LoS Model*

Let $p_n^{BS} = [x_n^{BS}, y_n^{BS}, z_n^{BS}] \in \mathbb{R}^3$ denote the coordinates of the $n^{th}$ BSs, where $n \in \mathcal{N} = \{1, ..., N\}$ and $N$ is the number of the BSs installed in the geographical area of interest; $p^{MT}(\tau) = [x^{MT}(\tau), y^{MT}(\tau)]$ denotes the MT (vehicle) position at time $\tau$; $p_m(\tau) = [x_m(\tau), y_m(\tau), z_m(\tau)] \in \mathbb{R}^3$ denote the coordinates of the $m^{th}$ unit cell of the RIS at time $\tau$, where $m \in \mathcal{M} = \{1, ..., M\}$ and $M$ is the number of the RIS elements. Since the RIS is fixed to the bottom of the UAV (facing to the ground), the coordinates of the RIS element can be calculated by the UAV coordinates and RIS/unit cell dimensions.

Let $\vec{\Delta}_{nm}^{BSRU}(\tau) = p_m(\tau) - p_n^{BS}$, $\vec{\Delta}_m^{RUMT}(\tau) = P^{MT} - p_m(\tau)$, and $\vec{\Delta}_n^{BSMT}(\tau) = P^{MT} - p_n^{BS}$ denote the vector from the $n^{th}$ BS to the $m^{th}$ unit cell of the RIS (i.e., the BS-RISoUAV link), the vector from the $m^{th}$ unit cell of the RIS to the MT (i.e., the RISoUAV-MT link), and the vector from the $n^{th}$ BS to the MT (i.e., the BS-MT link), respectively. For instance, $\vec{\Delta}_{nm}^{BSRU}(\tau) = \langle x_m(\tau) - x_n^{BS}, y_m(\tau) - y_n^{BS}, z_m(\tau) - z_n^{BS}\rangle$, or alternatively, $\vec{\Delta}_{nm}^{BSRU}(\tau) = (x_m(\tau) - x_n^{BS})\hat{\mathbf{i}} + (y_m(\tau) - y_n^{BS})\hat{\mathbf{j}} + (z_m(\tau) - z_n^{BS})\hat{\mathbf{k}}$, where $\hat{\mathbf{i}}, \hat{\mathbf{j}}$, and $\hat{\mathbf{k}}$ are the vectors of the unit length along the $x, y,$ and $z$ axes, respectively. The vector lengths, i.e., $|\vec{\Delta}_{nm}^{BSRU}(\tau)|, |\vec{\Delta}_m^{RUMT}(\tau)|$, and $|\vec{\Delta}_n^{BSMT}(\tau)|$ denote the BS-RISoUAV, RISoUAV-MT, and BS-MT distances, respectively. For instance:

$$|\vec{\Delta}_{nm}^{BSRU}(\tau)| \\ = \sqrt{(x_m(\tau) - x_n^{BS})^2 + (y_m(\tau) - y_n^{BS})^2 + (z_m(\tau) - z_n^{BS})^2} \quad (3)$$

The normalized vector of RISoUAV (on the xy-plane) is given as: $\vec{\Delta}^{RU} = \cos(\theta(\tau))\hat{\mathbf{i}} + \sin(\theta(\tau))\hat{\mathbf{j}}$, where $\theta$ is the heading (yaw) angle in (1). Suppose that $\Omega \in \mathbb{R}^3$ is a set of coordinates that belong to the 3D map of the dense urban area. Let $p_\Omega^{3D} = \{[x_{ij}^{3D}, y_{ij}^{3D}, z_{ij}^{3D}] | \forall \ i \in \{1, ..., \mathbb{P}_j\}, j \in \{1, ..., \mathbb{B}\} \in \Omega\}$, where $\mathbb{P}_j$ denotes the number of the Cartesian coordinates of the $j^{th}$ building and $\mathbb{B}$ is the number of buildings. The LoS for vectors $\vec{\Delta}_{nm}^{BSRU}(\tau), \vec{\Delta}_m^{RUMT}(\tau)$ and $\vec{\Delta}_n^{BSMT}(\tau)$ are modeled as:

$$\text{LoS}_{nm}^{BSRU}, \text{LoS}_m^{RUMT}, \text{LoS}_n^{BSMT} \\ = \begin{cases} 1, & if \ \vec{\Delta}_{nm}^{BSRU}(\tau), \vec{\Delta}_m^{RUMT}(\tau), \vec{\Delta}_n^{BSMT}(\tau) \cap p_\Omega^{3D} = \emptyset \\ 0, & if \ \vec{\Delta}_{nm}^{BSRU}(\tau), \vec{\Delta}_m^{RUMT}(\tau), \vec{\Delta}_n^{BSMT}(\tau) \cap p_\Omega^{3D} \neq \emptyset \end{cases} \quad (4)$$

which are treated as the LoS constraints in the RISoUAV navigation problem.

For illustration convenience, we assume that $p_m(\tau) = p(\tau), \forall \ m \in \mathcal{M}$. Fig. 2 shows the RISoUAV position and motion in the Cartesian coordinate system.

*C. Channel State Information (CSI)*

The RIS comprises a uniform linear array (ULA) of reflective elements as a specular reflector [5], and the phase shift of each element is controlled via an embedded controller at the UAV. The phase-shift matrix of the RIS is $\Theta(\tau) = \text{diag}\{e^{j\vartheta_1(\tau)}, e^{j\vartheta_2(\tau)}, ..., e^{j\vartheta_M(\tau)}\}$, where $\text{diag}(.)$ denotes a diagonal matrix and $\vartheta_i(\tau) \in [0, 2\pi), i \in \mathcal{M}$ is the phase-shift of the $i^{th}$ RIS element at time $\tau$. Here, it is worth noting that since RISoUAV chases a fast-moving vehicle, the compatibility of the response time of the RIS controller with RISoUAV motion must be considered in the optimization of the trajectory and speed which is the authors' future work.

Regarding the BS-MT link, since the LoS link can be randomly blocked by building and foliage, the Rayleigh channel fading model is utilized for taking the existing scatters. The BS-MT channel gain is given by:

$$g_{BSMT}(\tau) = \left(\sqrt{\rho |\vec{\Delta}_n^{BSMT}(\tau)|^{-\gamma}}\right)\tilde{g}, \quad (5)$$

where $\rho$ is the path loss at the reference distance, e.g., (1 m) [21], $\gamma \geq 2$ is the path loss exponent, and $\tilde{g}$ is a random scattering component modeled by a zero-mean and unit-variance circularly symmetric complex Gaussian (CSCG) random variable.

Since the RISoUAV trajectory is designed to secure the LoS link, the free-space path loss channel model is assumed in the BS-RISoUAV and RISoUAV-MT links. The channel gains are modeled as:

$$g_{BSRU}(\tau) \\ = \sqrt{\rho |\vec{\Delta}_n^{BSRU}(\tau)|^{-\gamma}} \left[1, e^{-j\frac{2\pi}{\lambda}d\phi_{BSRU}(\tau)}, ..., e^{-j\frac{2\pi(M-1)}{\lambda}d\phi_{BSRU}(\tau)}\right]^T \quad (6)$$

where $\phi_{BSRU}(\tau)$ is the cosine of the angle-of-arrival (AoA) of the incident signal from the BS to the RISoUAV at time $\tau$, $d$ is the antenna spacing, and $\lambda$ is the carrier wavelength. Similarly, the channel gain of the RISoUAV-MT link is given by:

$$g_{RUMT}(\tau) = \sqrt{\rho |\vec{\Delta}_n^{RUMT}(\tau)|^{-\gamma}} \left[1, e^{-j\frac{2\pi}{\lambda}d\phi_{RUMT}(\tau)}, \ldots, e^{-j\frac{2\pi(M-1)}{\lambda}d\phi_{RUMT}(\tau)}\right]^T \quad (7)$$

where $\phi_{RUMT}(\tau)$ is the cosine of the angle-of-departure (AoD) of the incident signal from the RIS to the MT.

The signal-to-noise ratio (SNR) at the MT through the BS-RISoUAV-MT and direct BS-MT links is obtained as:

$$\Upsilon_{BS-RU-MT}(\tau) = \frac{P_{BS}(\tau)|g_{BSMT}(\tau) + g_{RUMT}^H(\tau)\Theta(\tau)g_{BSRU}(\tau)|^2}{\sigma^2} \quad (8)$$

where $\Upsilon_{BS-RU-MT}(\tau)$ is the SNR at time $\tau$, $P_{BS}(\tau)$ is the power transmitted by the BS at time $\tau$, and $\sigma^2$ is the noise variance. Finally, the achievable rate at the MT is obtained as:

$$R_{BS-RU-MT}(\tau) = \log_2\bigl(1 + \Upsilon_{BS-RU-MT}(\tau)\bigr) \quad (9)$$

### D. Problem Formulation

We aim to achieve three objectives for the RISoUAV mission:

1) Minimizing the energy consumption of the UAV given by (2), subject to the constraints given in (1):

$$\min_{v(t),\omega(t),u(t)} \{E_{RISoUAV}\} \quad (10)$$

This also ensures the path with the minimum length is selected for the RISoUAV. This is helpful to avoid violating the speed limits by the UAV (because the UAV must follow MT, which may move fast).

2) Securing uninterrupted BS-RIS-MT LoS link and SNR constraint. Based on the LoS model in (4) and SNR in (8), the LoS and SNR constraints are defined as:

$$\bigl(\vec{\Delta}_n^{BSRU}(\tau) \cup \vec{\Delta}^{RUMT}\bigr) \cap p_\Omega^{3D} = \emptyset$$

$$\bigl(-\vec{\Delta}_n^{BSRU}(\tau), \vec{\Delta}^{RU}(\tau)\bigr) > 0 \ \& \ \bigl(\vec{\Delta}^{RUMT}(\tau), \vec{\Delta}^{RU}(\tau)\bigr) > 0 \quad (11)$$

$$SRN \geq SNR_{min}$$

where $(.,.)$ denotes the inner product of two vectors. The second line of (11) ensures that both the BS and the MT are on the same side of the RISoUAV. Since we assume that the RIS faces the ground, this constraint does not apply to the BS-RISoUAV-MT application.

3) Passive beamforming and maximizing the achievable rate at the MT through the BS-RIS-MT link. From (8), we have $\Upsilon_{BS-RU-MT}(\tau) \geq 0$. Based on (9), we have $R_{BS-RU-MT}(\tau) \propto \Upsilon_{BS-RU-MT}(\tau)$. Regarding the performance of the BS-RIS-MT link, we consider the following objective function for optimizing the channel performance:

$$\max_{p(t),\theta(t),\Theta} \{\Psi\} \quad (12)$$

where

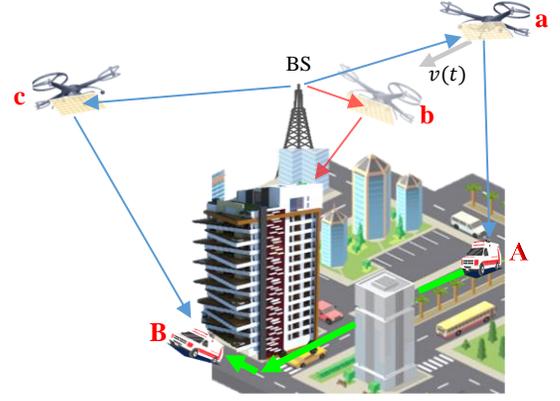

Fig. 3. Challenge in 3D trajectory design and the RISoUAV navigation considering LoS: suppose MT is at position "A" and LoS link is on while RISoUAV is at position "a". When MT reaches to position B and the RISoUAV reaches to position "b", the LoS is immediately lost. Thus, the RISoUAV should immediately move from position "b" to position "c" with a very fast speed that may violate speed limit, otherwise the communication link is interrupted.

$$\Psi = \int_{t_0}^{t_0+T} g_{RUMT}^H(\tau)\Theta(\tau)g_{BSRU}(\tau)\,d\tau$$

$$\Rightarrow \Psi = \int_{t_0}^{t_0+T} \frac{\rho}{\sqrt{(|\vec{\Delta}^{RUMT}[\epsilon]||\vec{\Delta}_n^{BSRU}[\epsilon]|)^\gamma}} \times$$

$$\sum_{m=1}^M e^{j\left(\vartheta_m(\tau) + \frac{2\pi}{\lambda}d(m-1)(\phi_{RUMT}(\tau) - \phi_{BSRU}(\tau))\right)}\,d\tau;$$

$$\phi_{RUMT}(\tau) = \frac{\sqrt{(x^{MT}(\tau)-x(\tau))^2 + (y^{MT}(\tau)-y(\tau))^2}}{|\vec{\Delta}^{RUMT}(\tau)|};$$

$$\phi_{BSRU}(\tau) = \frac{\sqrt{(x(\tau)-x_n^{BS})^2 + (y(\tau)-y_n^{BS})^2}}{|\vec{\Delta}_n^{BSRU}(\tau)|}.$$

## III. THE PROPOSED SOLUTION FOR RISOUAV NAVIGATION

The objectives given in (10)-(12) are nonlinear non-convex optimization problems. Besides, there are other issues in solving the problem which are explained in the next section.

### A. The RISoUAV Navigation problem

The MT map route $\Re$, from the start position of the MT to its goal, is known, as it can be yielded by maps Apps. The MT communicates the map route with the UAV. The UAV must hold the instant (BS-RIS-MT) LoS link while satisfying the speed and SNR constraint through an energy-efficient path with the optimum channel performance. To enable determining the valid/optimal trajectory of the RISoUAV, the possible solution is to discretize the route into $K$ points with relatively short distances. For every point, the optimal RISoUAV's 3D position is obtained regarding the energy efficiency, LoS, and channel performance. Then connecting the optimal points gives rise to the optimal trajectory. However, there is an issue with this method, which is explained as follows.

Both distances between the BS and RISoUAV, and between the RISoUAV and MT affect the performance of the BS-RIS-MT communication channel. Besides, the AoA and AoD affect the channel gain.

Table I. Stage 1. Energy-efficient and LoS-secured tube path

| # | Expression | Description |
|---|---|---|
| 1. | $\mathfrak{R}: \{\mathfrak{R}[k] \mid \forall k \in \mathcal{K} = \{0, 1, ..., K\}\}$ | Divide $\mathfrak{R}$ (MT route) to $K$ points |
| 2. | $\mathcal{T}: \{\mathcal{T}[k] \mid \forall k \in \mathcal{K} = \{0, 1, ..., K\}\}$ and $T = \sum_{k=1}^{K} \mathcal{T}[k]$ | MT travel time |
| 3. | $\mathcal{Q}[k, q_k] \leftarrow Random\,([p[k, q_k], \theta[k, q_k]])$, $\forall q_k \in \mathbb{Q}_k = \{1, ..., Q_k\}, \forall k \in \mathcal{K}$ | Random positions at each $k$ |
| 4. | $\mathcal{S}[k, q_k] \leftarrow (\mathcal{Q}[k, q_k], r)$ $\forall q_k \in \mathbb{Q}_k, \forall k \in \mathcal{K}$ | Spheres |
| 5. | $(\vec{\Delta}_n^{BSRU}[k, q_k] \cup \vec{\Delta}^{RUMT}[k, q_k]) \cap p_\Omega^{3D} = \emptyset$ $\forall q_k \in \mathbb{Q}_k, \forall k \in \mathcal{K}$ $SRN \geq SNR_{min}$ | Check LoS and $SNR_{min}$ constraints in (11) |
| 6. | $U[k, q_{k-1}^k] = \dfrac{\mathcal{Q}[k, q_k] - \mathcal{Q}[k-1, q_{k-1}]}{\mathcal{T}[k]}$ $\forall q_k \in \mathbb{Q}_k, \forall q_{k-1} \in \mathbb{Q}_{k-1}, \forall k \in \mathcal{K}$ | UAV speed (13) |
| 7. | $|U[k, q_{k-1}^k]| \leq U_{max}$ $\forall q_{k-1} \in \mathbb{Q}_{k-1}, \forall k \in \mathcal{K}$ | Speed constraints (1) |
| 8. | $\Xi_\aleph(\mathcal{Q}[k, q_k], \forall k \in \mathcal{K})$ $= \{\mathcal{Q}[1, q_1], ..., \mathcal{Q}[k, q_k], ..., \mathcal{Q}[K, q_K]\}$ | $\aleph$ prospective trajectories with valid speeds |
| 9. | $\dot{U}[k, q_{k-1}^k] = \dfrac{U[k, q_{k-1}^k] - U[k-1, q_{k-2}^{k-1}]}{\mathcal{T}[k]}$ | UAV acceleration |
| 10. | $|\dot{U}[k, q_{k-1}^k]| \leq \dot{U}_{max}$ $\forall q_{k-1} \in \mathbb{Q}_{k-1}, \forall k \in \mathcal{K}$ | Acceleration constraints (14) |
| 11. | $E_{RISoUAV}^{\Xi_g} = \sum_K \alpha_1 U^{\Xi_g}[k] + \alpha_2 \dot{U}^{\Xi_g}[k]$ | The energy consumption of $\Xi_g$ (15) |
| 12. | $\Xi_g(\mathcal{Q}[k, q_k], \forall k \in \mathcal{K}): \min E_{RISoUAV}^{\Xi_{1-\aleph}}$ | prospective trajectory with minimum energy |
| 13. | $\Phi(S^{\Xi_g}[k, q_k], U^{\Xi_g}[k] \forall k \in \mathcal{K})$ $= \left\{\begin{array}{l}(S^{\Xi_g}[1, q_1], U^{\Xi_g}[1]), ..., \\ (S^{\Xi_g}[K, q_K], U^{\Xi_g}[K])\end{array}\right\}$ | Tube path and associated input (speed) |

So, the RISoUAV may have a non-zero horizontal distance from the MT while taking the BS location into account. Also, the RISoUAV is controlling its speed and direction to make sure the BS-RIS-MT LoS link is continuously held with an acceptable achievable rate. However, in some situations, depending on the arrangement of buildings, the LoS might immediately lose (e.g., if the MT changes its direction to a crossing street in an intersection) and the RISoUAV must take immediate action to take a new position with a possible LoS. It may violate the speed constraints. This issue is illustrated in Fig. 3. To solve this problem, a receding horizon (with some receding slots) can be taken, instead of making decisions for every single forthcoming slot.

However, the use of receding horizons imposes a computational burden for solving (10)-(12), as the receding horizon (i.e., the number of selected forthcoming points) should be large enough since the MT is moving fast whereas the distances between the sampling points should be short. To address this issue, we consider a two-stage trajectory optimization for RISoUAV navigation, as described in the next section.

### B. A Two-Stage Optimal Navigation Model for RISoUAV

*Stage 1. Energy-efficient and LoS-secured tube path*

At this stage, the RISoUAV's random trajectories are determined by utilizing the RRT algorithm. Then the MPC method [22] is used to achieve the optimal trajectory with the minimum UAV energy consumption, which satisfies the problem constraints, i.e., speed/acceleration limits and LoS constraints.

To this end, the MT route $\mathfrak{R}$ is divided into $K$ points with relatively large distances. Then the map route is modeled by $\mathfrak{R}: \{\mathfrak{R}[k] \mid \forall k \in \mathcal{K} = \{0, 1, ..., K\}\}$ that includes the start, middle, and endpoints of streets (direct routes) in $\mathcal{K}$. Let $\mathcal{T}[k]$ be the time indicator (i.e., the time period) that the MT takes to move from $\mathfrak{R}[k-1]$ to $\mathfrak{R}[k]$ (i.e., the position change of the MT) and is available based on the MT speed, so that $T = \sum_{k=1}^{K} \mathcal{T}[k]$.

For $\mathfrak{R}[k], \forall k \in \mathcal{K}$, some random positions are assigned to the RISoUAV as given by:
$\mathcal{Q}[k, q_k] = Random\,([p[k, q_k], \theta[k, q_k]]), \forall q_k \in \mathbb{Q}_k = \{1, ..., Q_k\}$, where $Q_k$ is the number of random positions for the $k^{th}$ point. We define $\mathcal{S}[k, q_k]$ as the sphere which includes $\mathcal{Q}[k, q_k]$ as the center with radius $r$. Those of $\mathcal{Q}[k, q_k]$ for which the LoS and $SRN$ prerequisite in (11)-(12) are not satisfied and/or the associated sphere overlaps with the pre-specified spheres are omitted. Also, there should be a valid direct path between random positions of point $\mathfrak{R}[k]$ and allocated random positions of point $\mathfrak{R}[k-1]$.

When the MT moves from $\mathfrak{R}[k-1]$ to $\mathfrak{R}[k]$, we yield the UAV input (average speed) associated with the point-to-point prospective trajectory starting from $\mathcal{Q}[k-1, q_{k-1}]$ to $\mathcal{Q}[k, q_k]$, which is defined as

$U[k, q_{k-1}^k] = [v_x[k, q_{k-1}^k], v_y[k, q_{k-1}^k], u[k, q_{k-1}^k], \omega[k, q_{k-1}^k]]$,

where $U[k, q_{k-1}^k]$ is obtained for $\forall q_k \in \mathbb{Q}_k, \forall q_{k-1} \in \mathbb{Q}_{k-1}$ and by injecting $\mathcal{Q}[k-1, q_{k-1}], \forall q_{k-1} \in \mathbb{Q}_{k-1}$ and $\mathcal{Q}[k, q_k], \forall q_k \in \mathbb{Q}_k$ into (13) which is discretized version of (1), as given by

$$U[k, q_{k-1}^k] = \dfrac{\mathcal{Q}[k, q_k] - \mathcal{Q}[k-1, q_{k-1}]}{\mathcal{T}[k]}, \quad (13)$$
$\forall k \in \mathcal{K}, \forall q_k \in \mathbb{Q}_k, \forall q_{k-1} \in \mathbb{Q}_{k-1}$

So we have $Q_k Q_{k-1}$ number of possible RISoUAV point-to-point trajectories from $\mathfrak{R}[k-1]$ to $\mathfrak{R}[k]$.

We count $\aleph$ number of prospective trajectories for $\mathfrak{R}$ starting from $\mathfrak{R}[0]$ to $\mathfrak{R}[K]$ as:

$\Xi_{1:\aleph}(\mathcal{Q}[k, q_k], \forall k \in \mathcal{K}) = \{\mathcal{Q}[1, q_1], ..., \mathcal{Q}[k, q_k], ..., \mathcal{Q}[K, q_K]\}$

by considering all possible connections of point-to-point $\mathcal{Q}[k-1, q_{k-1}]$ to $\mathcal{Q}[k, q_k] \forall q_k \in \mathbb{Q}_k, \forall k \in \mathcal{K}$ paths. Totally $\aleph = \prod_{k=1}^{K-1} Q_k Q_{k+1}$ is the number of prospective trajectories through the route $\mathfrak{R}$ from $\mathfrak{R}[0]$ to $\mathfrak{R}[K]$. Nevertheless, the practical numbers of possible RRT branches are far smaller than $\aleph$ because the corresponding branch of the tree is discarded every time if the speed/acceleration constraints are not satisfied. The speed constraints can be found in (1) and acceleration constraints are given as:

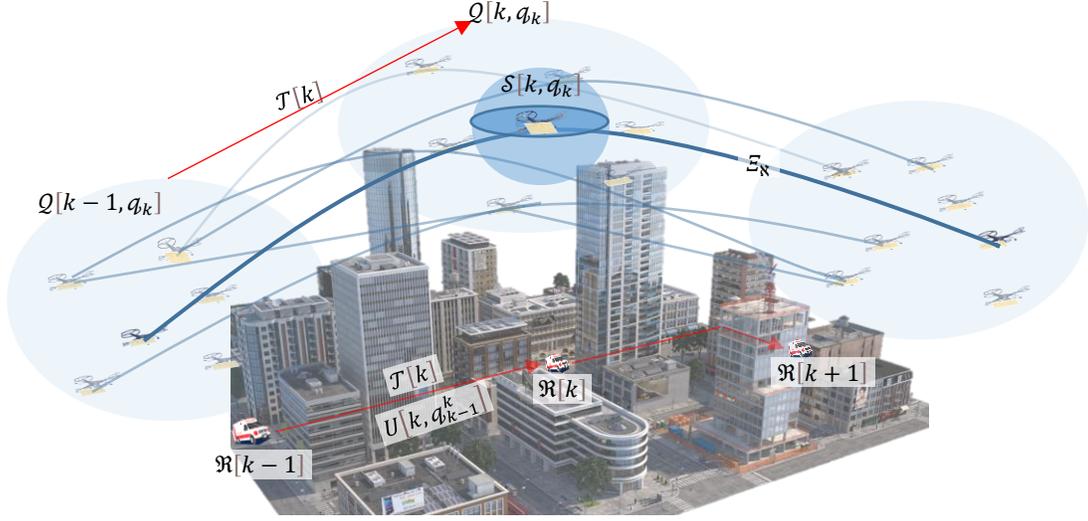

Fig. 4. RISoUAV navigation through utilizing the rapidly exploring random tree (RRT) algorithm and MPC.

Table II. Stage 2. RISoUAV trajectory and passive beamforming to get optimum channel performance

| | | |
|---|---|---|
| 1. | $\boldsymbol{\Phi}[k]$: $\{\boldsymbol{\Phi}[k,\epsilon] \mid \forall \epsilon \in \mathbb{E} = \{1, \dots, \mathcal{E}\}\}$ <br> $\mathcal{S}[k,\epsilon]$: $\boldsymbol{\Phi}[k,\epsilon] \mid \forall \epsilon \in \mathbb{E}$ | Divide $\boldsymbol{\Phi}[k]$ to $\mathcal{E}$ slots and find the relevant sphere |
| 2. | $p[k,\epsilon,\mathcal{b}_\epsilon] \leftarrow Rand.\, p[k,\epsilon] \in \mathcal{S}[k,\epsilon], \forall \epsilon \in \mathbb{E}$ <br> $\forall \mathcal{b}_\epsilon \in \mathcal{B}_\epsilon = \{1, \dots, B_\epsilon\}$ | Random positions the $\mathcal{S}[k,\epsilon]$ in $\boldsymbol{\Phi}[k,\epsilon]$ |
| 3. | $\left(\vec{\Delta}_n^{BSRU}[\epsilon,\mathcal{b}_\epsilon] \cup \vec{\Delta}^{RUMT}[\epsilon,\mathcal{b}_\epsilon]\right) \cap p_\Omega^{3D} = \emptyset$ <br> $\forall \epsilon \in \mathbb{E}, \forall \mathcal{b}_\epsilon \in \mathcal{B}_\epsilon = \{1, \dots, B_\epsilon\}$ <br> $SRN \geq SNR_{min}$ | Check LoS and $SNR_{min}$ constraints in (11) |
| 4. | $p[k,\epsilon,\mathcal{b}] \leftarrow \phi_{BSRU}[k,\epsilon,\mathcal{b}], \phi_{RUMT}[k,\epsilon,\mathcal{b}]$ | Find AoA and AoD for each position |
| 5. | $\vartheta_m[k,\epsilon,\mathcal{b}] = \frac{2\pi d(m-1)}{\lambda}(\phi_{BSRU}[k,\epsilon,\mathcal{b}] - \phi_{RUMT}[k,\epsilon,\mathcal{b}]) + \varpi$ | Determine phase shift matrix for each random position |
| 6. | $\Xi^*_{1:\aleph^*}[k]$: $\{p[k,\epsilon,\mathcal{b}] \mid \forall \epsilon \in \mathbb{E}, \forall \mathcal{b} \in \mathcal{B}_\epsilon\}$ <br> $= \{p[k,1,\mathcal{b}_1], \dots, p[k,\mathcal{E},\mathcal{b}_\mathcal{E}]\}$ | $\aleph^*$ prospective trajectories |
| 7. | $f(\Xi^*[k]) = \sum_\epsilon^\mathcal{E} \|\vec{\Delta}^{RUMT}[\epsilon,\mathcal{b}_\mathcal{E}]\|\|\vec{\Delta}_n^{BSRU}[\epsilon,\mathcal{b}_\mathcal{E}]\|$ <br> $1:\aleph^*$ <br> $p[k,\epsilon,\mathcal{b}] \in \mathcal{S}[k,\epsilon], \forall \epsilon \in \mathbb{E}, \forall \mathcal{b} \in \mathcal{B}_\epsilon$ | Calculate (19) |
| 8. | Select $\Xi^*[k]$ for $\aleph^{*th}$ trajectory with $\min f(\Xi^*[k])$ | The final trajectory with most fitted (19) |
| 9. | $\Xi^* = \{\Xi^*[1], \dots, \Xi^*[K]\}$ | Final path |

$$\left|\frac{v[k,q_{k-1}^k] - v[k-1,q_{k-2}^{k-1}]}{\mathcal{T}[k]}\right| \leq \dot{v}_{max}$$
$$\left|\frac{u[k,q_{k-1}^k] - u[k-1,q_{k-2}^{k-1}]}{\mathcal{T}[k]}\right| \leq \dot{u}_{max} \quad (14)$$
$$\left|\frac{\omega[k,q_{k-1}^k] - \omega[k-1,q_{k-2}^{k-1}]}{\mathcal{T}[k]}\right| \leq \dot{\omega}_{max}$$

where $\dot{v}_{max}$, $\dot{u}_{max}$ and $\dot{\omega}_{max}$ denote horizontal, vertical, and angular acceleration limits, respectively. In this way, the navigation concern illustrated in Fig. 3 is obviated.

For each of the remaining branches (for each prospective trajectory) the energy cost function is solved using (15) which is developed by discretizing (2):

$$\begin{aligned}E_{RISoUAV}^{\Xi_g} = \sum_{k=1}^{K} &\alpha_1\big(v_x[k,q_{k-1}^k] + v_y[k,q_{k-1}^k] + |u[k,q_{k-1}^k]| \\ &+ |\omega[k,q_{k-1}^k]|\big) \\ &+ \alpha_2 \frac{|v_x[k,q_{k-1}^k] - v_x[k-1,q_{k-2}^{k-1}]|}{\mathcal{T}[k]} \\ &+ \alpha_2 \frac{|v_y[k,q_{k-1}^k] - v_y[k-1,q_{k-2}^{k-1}]|}{\mathcal{T}[k]} \\ &+ \alpha_2 \frac{|u[k,q_{k-1}^k] - u[k-1,q_{k-2}^{k-1}]|}{\mathcal{T}[k]} \\ &+ \alpha_2 \frac{|\omega[k,q_{k-1}^k] - \omega[k-1,q_{k-2}^{k-1}]|}{\mathcal{T}[k]}\end{aligned} \quad (15)$$

where $E_{RISoUAV}^{\Xi_g}$ denotes the energy consumption of the UAV through the $g^{th}$ prospective trajectory ($\Xi$).

The trajectory with the minimum energy is adopted as the final path. Finally, the RISoUAV tube path $\Phi$ is achieved by connecting the spheres associated with the selected trajectory $\Xi_g$, as given by

$$\Phi\big(\mathcal{S}^{\Xi_g}[k,q_k], \forall\, k \in \mathcal{K}\big) = \{\mathcal{S}^{\Xi_g}[1,q_1], \dots, \mathcal{S}^{\Xi_g}[k,q_k], \dots, \mathcal{S}^{\Xi_g}[K,q_K]\}$$

The RISoUAV navigation model for minimizing the energy consumption that satisfies flight constraints is summarized in Table I, as shown in Fig. 4.

In the next section, the exact RISoUAV trajectory (i.e., $p(t), \theta(t)$) is obtained through the tube path $\Phi$ by optimizing channel performance, i.e., achievable rate and phase-shift matrix. In this light, a compromise between problem scalability and solution feasibility should be made for the selection of the number of random positions and sphere radius $r$. The larger $r$ the fewer possible random positions, and thus the computation burden of the problem is reduced. So a large $r$ can be selected if the computation cost of the UAV navigation board is limited. However, the feasibility of the solution can be invalid as the UAV speed limit might be violated.

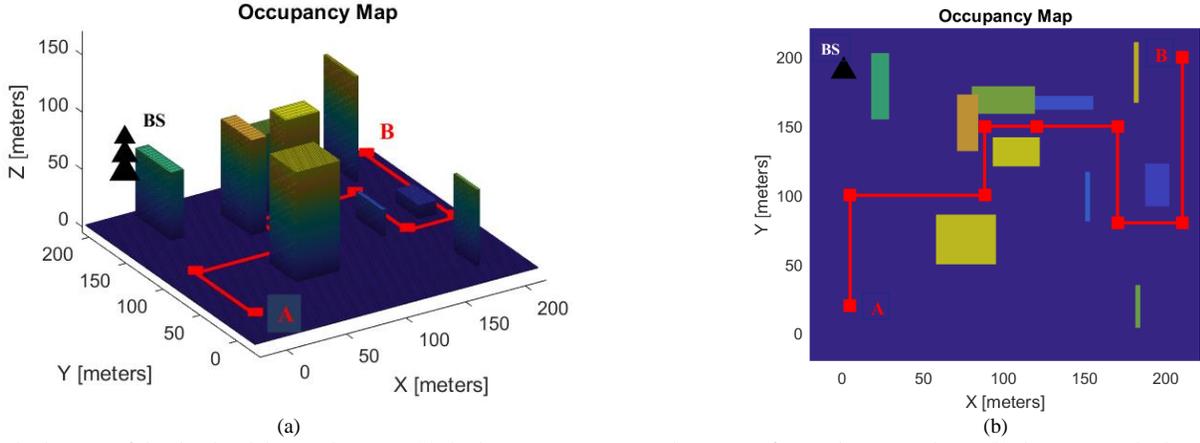

Fig. 5. The 3D map of the simulated dense urban area: (a) the 3D occupancy map and MT route from point A to point B; (b) the MT route in the XY plane. The MT map route is divided into 8 slots by 9 points: $\Re[k]| \ \forall \ k = 1, \ldots, 9$, which are shown by red squares.

Besides, the sphere radius $r$ should not be too large, as the speed variations of the final solution (through stage 2) can be tolerated. On the other hand, a small $r$ increases the computational burden but enhances the accuracy and thus the reliability of the solution.

*Stage 2. RISoUAV trajectory and passive beamforming to achieve optimum channel performance*

After achieving the secure tube path, the channel performance is considered for obtaining the optimal RISoUAV trajectory through the tube path. To this end, optimizing the objective function in (12), i.e., maximizing the achievable rate at the MT through the BS-RISoUAV-MT link, is considered. However, the optimization problem in (12) is non-convex concerning the RISoUAV trajectory variables, i.e, $Q(t) = (p(t), \theta(t))$ and phase-shift matrix of the RIS ($\Theta(t)$).

We solve the optimization problem in (12) by using tube-based receding horizon predictive control. The tube-based method limits possible solutions to predefined/valid boundaries, which helps to reduce the computational burden. In this regard, we divide (discretize) the tube path associated with $\Re[k-1]$ and $\Re[k]$, i.e., $\Phi[k]$, into $\mathcal{E}$ slots as $\Phi[k,\epsilon] \ \forall \ \epsilon \in \mathbb{E} = \{1, \ldots, \mathcal{E}\}$ where $\mathcal{E}$ is considered as the receding horizon for maximizing achievable rate through the communication channel for the upcoming point-to-point trajectory. We assume the speed/acceleration constraints are satisfied through the tube path and thus are not considered in this stage, to reduce the computational burden. Therefore, more slots (larger $\mathcal{E}$) can be adopted to increase the accuracy of the model.

Thus, the achievable rate over the point-to-point trajectory (and accordingly throughout the obtained trajectory) is increased. However, the LoS and SNR constraints should be satisfied. Since the RISoUAV navigation is optimized in the first stage and the RIS faces to the ground, the channel performance is independent of $\theta[k, \epsilon]$ (i.e., the heading of the UAV with respect to the x-axis), thanks to the intelligent phase-shift matrix[1] and passive beamforming. Thus, $\theta[k, \epsilon]$ is not considered as an optimizing variable and can be achieved based on the UAV positions at two consecutive slots (i.e., $\Phi[k, \epsilon-1]$ and $\Phi[k, \epsilon]$). Thus, (12) is updated for $\Phi[k]$ as:

$$\Psi(p[k,\epsilon], \Theta[k,\epsilon], \forall \epsilon \in \mathbb{E})$$
$$= \rho \sum_{\epsilon=1}^{\mathcal{E}} \sum_{m=1}^{M} \frac{e^{j\left(\vartheta_m[k,\epsilon] + \frac{2\pi}{\lambda}d(m-1)(\phi_{RUMT}[k,\epsilon] - \phi_{BSRU}[k,\epsilon])\right)}}{\sqrt{\left(|\vec{\Delta}^{RUMT}[k,\epsilon]||\vec{\Delta}_n^{BSRU}[k,\epsilon]|\right)^\gamma}} \quad (16)$$

$$p[k,\epsilon] \in \Phi[k,\epsilon], \vartheta_m[\epsilon] \in [0, 2\pi] \forall \epsilon \in \mathbb{E}$$

To maximize the achievable rate, we can regulate the phase-shift of the RIS elements, i.e., $\vartheta_m \forall m \in \mathcal{M}$. As a result, the associated received energy of all signals MT accumulated coherently. To achieve this, we set:

$$\vartheta_m[k,\epsilon] = \frac{2\pi d(m-1)}{\lambda}(\phi_{BSRU}[k,\epsilon] - \phi_{RUMT}[k,\epsilon]) + \varpi \quad \forall \epsilon, m \quad (17)$$

where $\varpi \in [0, 2\pi]$. Thus (16) is updated as:

$$\Psi(p[k,\epsilon], \forall \epsilon \in \mathbb{E}) = \rho \sum_{\epsilon=1}^{\mathcal{E}} \frac{M}{\sqrt{\left(|\vec{\Delta}^{RUMT}[k,\epsilon]||\vec{\Delta}_n^{BSRU}[k,\epsilon]|\right)^\gamma}} \quad (18)$$

$$p[k,\epsilon] \in \Phi[k], \forall \epsilon \in \mathbb{E}$$

Then the problem is transformed to minimizing the denominator in (18) that can be written as:

$$\min_{p[k,\epsilon]} \sum_{\epsilon}^{\mathcal{E}} \left|\vec{\Delta}^{RUMT}[k,\epsilon]\right| \left|\vec{\Delta}_n^{BSRU}[k,\epsilon]\right|$$
$$\text{s.t.} \ p[k,\epsilon] \in \mathcal{S}[k,\epsilon] \forall \epsilon \in \mathbb{E} \quad (19)$$

where $\mathcal{S}[k,\epsilon]$ is the sphere associated with the $\epsilon^{th}$ slot in $\Phi[k]$. The optimization problem is to find a set of positions $\{p[k,\epsilon] \in \mathcal{S}[k,\epsilon]| \ \forall \epsilon \in \mathbb{E}\} \in \Phi[k]$ for the RISoUAV so that (19) is minimized through tube path $\Phi[k]$, which is nonlinear convex. To solve (19) we use the RRT algorithm. Some random prospective positions are assigned to $\mathcal{S}[k,\epsilon], \forall \epsilon \in \mathbb{E}$. Then, we have $\{p[k,\epsilon,\ell_\epsilon]| \ \forall \ \ell_\epsilon \in \mathcal{B}_\epsilon = \{1, \ldots, B_\epsilon\}, \forall \epsilon \in \mathbb{E}\}$, where $B_\epsilon$ is the number of random positions for the $\epsilon^{th}$ slot. The trajectory, including $\{p[k,1,\ell_1], \ldots, p[k,\mathcal{E},\ell_\mathcal{E}]\}$ that satisfies (19), is selected as the final UAV navigation path.

---

[1] Here we assume that the response time of the intelligent phase-shift matrix (as an electronics-based element) is fast enough compared with UAV maneuver speed (as a mechanical machine). However, the UAV heading can be considered in the optimization process considering the response time and performance of the intelligent phase shift matrix of the RIS. This issue value more investigation which is considered as future works.

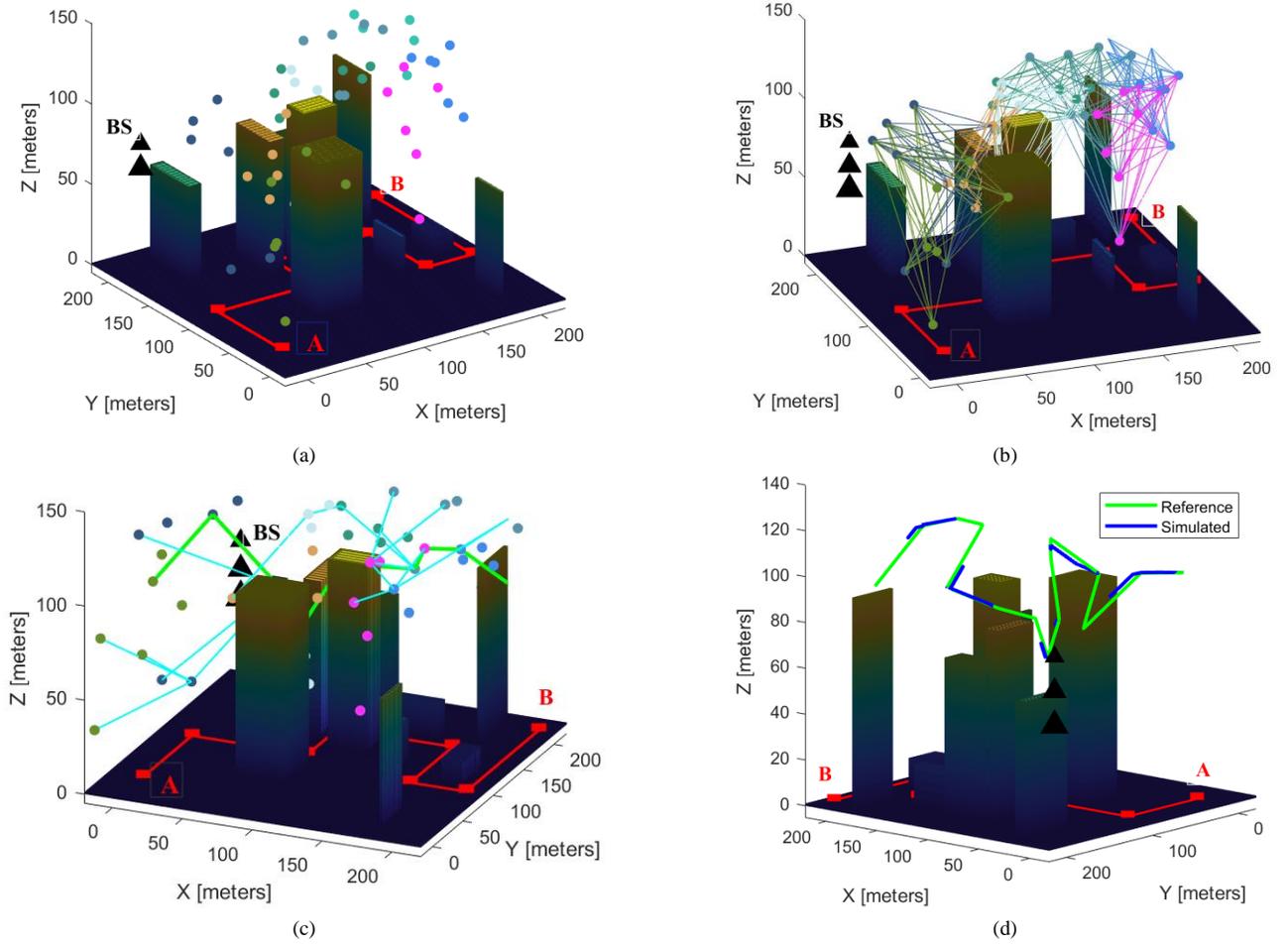

Fig. 6. Simulation results for the dense urban scenario: (a) stage 1, assigning 6 random positions $Q[k, q_k] \forall q_k = 1, ..., 5$ for each point $\Re[k]| \forall k = 1, ..., 9$, indicated positions with the same color are associated to the same route point; (b) all possible trajectories through random positions for obtaining speeds/accelerations; (c) $\aleph$ prospective trajectories with valid speeds/accelerations (shown by light blue lines) and selected trajectory with minimum energy consumption (shown by light green lines) which specifies the tube path; (d) the final optimum path through the second stage as the reference path (shown by light green lines) and simulating UAV navigation following the planned path (shown by blue lines).

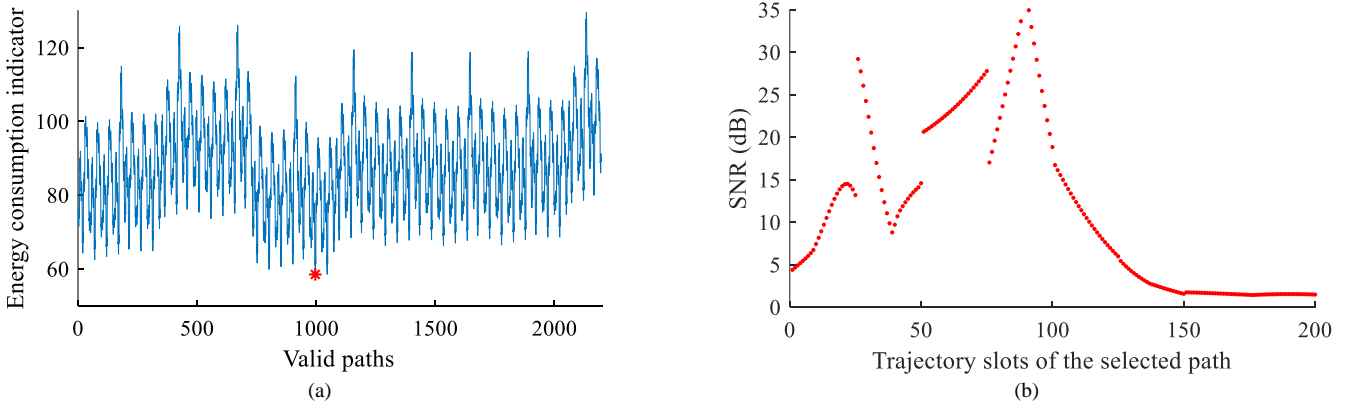

Fig. 7. Numerical result for the simulated scenario: (a) Energy consumption indicator for valid paths; (b) SNR for the trajectory slots of the selected path.

Stage 2 for finding the final trajectory of the RISoUAV with the optimal channel performance through the optimum energy-efficient tube path is summarized in Table II.

## IV. SIMULATION RESULTS

The RISoUAV navigation in a dense urban area is simulated in the Matlab platform to evaluate the effectiveness of the proposed method. The 3D occupancy map of the simulated scenario is presented in Fig. 5. The MT is moving from point A to point B through the MT route shown by the red lines. The occupancy map is validated and through the Matlab UAV toolbox (obstacle-free map locations are considered as valid states and occupied and unknown map locations are interpreted as invalid states). The system parameters are $P_{BS} = 30$ dBm, $\sigma^2 = -80$ dBm, $\rho = 10$ dBm, $\gamma = 2.5$, $\lambda = 10^{-2}$ m, $d=d = \frac{1}{2}\lambda$.

To find an energy-efficient and LoS-secured tube path for the RISoUAV to follow the MT, the A-B route is divided into 8 slots by choosing 9 points including the crosspoint of streets and middle of the long streets in dense areas. The MT travel time (the time indicator), based on which the RISoUAV navigation and speed/acceleration limits are determined is selected as

$\mathcal{T}[k] = \{8 \quad 8.3 \quad 5.2 \quad 4.1 \quad 4 \quad 7 \quad 4 \quad 12\}$ seconds.

For each point $\Re[k]| \forall k = 1, ..., 9$, 6 random positions $\mathcal{Q}[k, q_k] \forall q_k = 1, ..., 6$ are allocated, see Fig. 6(a), that satisfy the following constraints; $35\ m \leq z \leq 130\ m$, $r = 15\ m$, and maximum horizontal distance from point $k$ (i.e., $\Re[k]$) for allocated random positions is 50 m.

Also, the allocated random positions must satisfy the BS-RIS-MT LoS and SNR constraints and there must be a valid path between $\mathcal{Q}[k-1, q_{k-1}]$ and $\mathcal{Q}[k, q_k]$. Increasing (decreasing) the number of allocated random positions for each point (the sphere radius $r$) increases the chance of getting a valid optimum tube path, but imposes the computational burden.

The number of trajectories through the allocated random positions is very large (i.e., $(6 \times 6)^{9-1}$), see Fig. 6(b). However, after applying the RISoUAV speed/acceleration constraints ($V_{max} = 12\frac{m}{s}, U_{max} = 8\frac{m}{s}, W_{max} = \pi/6 \frac{rad}{slot}$), the number of valid prospective paths reduces to 2196, see Fig. 6(c). Then, the path with minimum energy consumption is selected by which the energy-efficient and LoS/SNR-secured tube path for the second stage is achieved, see Fig. 7(a). At the second stage, using RRT and based on the channel performance rule in (19) the final optimal trajectory is obtained, see Fig. 6(d). The number of assigned slots for each point-to-point path is selected 25 for the second stage (a total of 200 slots for the whole RISoUAV trajectory). The achieved SNRs for trajectory slots ($SNR_{min} = 1$) are indicated in Fig. 7(b)

*Discussion:* Based on the simulation results, illustrated in Fig. 6, the proposed method successfully discovers an energy-efficient and LoS-secured path for RISoUAV navigation to provide an instant LoS wireless communication channel for the MT. However, designing the system parameters (for modeling the problem and developing a solution) depends on the UAV and environmental characteristics. For instance, in terms of the computational burden, selecting the number of planning slots and sphere radius $r$ depend on the performance of the computational board. From the UAV flight perspective, the UAV maneuvers (speed/acceleration) limits, the maximum flight altitude for allocating potential positions for each route point (at stage 1), the height of buildings, density of the urban area, etc., impact the performance and convergence of the method. From the channel performance point of view, achieving an acceptable SNR (while satisfying the LoS and flight constraints) depend on the BSs allocations, power rating of the incident signals, number of RIS elements, the flight altitude, etc. In this light, in the simulated scenario, we have considered one BS to evaluate if the proposed method can find an appropriate navigation path for the RISoUAV to satisfy problem constraints. So, it can be seen that the achieved SNR at the last slots (where the MT is far from the BS and the UAV has to fly at a higher altitude to secure the LoS) is declining. Installing other BSs, strengthening the power rating of the communication network, or adopting larger RISs with more elements can be helpful but cost more. Thus, securing convergence of the method to a valid solution demands a compromise between system (level of) security requirements and costs based on environmental characteristics.

## V. Conclusion

The navigation of a UAV equipped with a mounted RIS was designed for providing an instant LoS communication channel for a mobile vehicle (an emergency ambulance) in the 5G (mmWave) wireless communication networks. To tackle the computational burden for achieving an optimal energy-efficient and LoS-secured path while considering the UAV flight constraints and BS-RIS-MT communication channel performance, a two-stage method was proposed. Stage 1 is executed through an offline process (while the RISoUAV is preparing for the trip) to find an LoS-secured tube path with minimum energy consumption that satisfies the RISoUAV maneuver (speed/acceleration) constraints. Stage 2 is executed during the flight for the next receding horizon aiming at maximizing the achievable rate through the communication channel.

The performance/adjustment of the phase shift matrix in accordance with the UAV maneuver has not been considered in this work and can be considered as future works. Also, uncertainty in MT motion due to traffic and other vehicles is an issue that can be considered to extend the work. Besides, the UAV navigation design when the 3D map is not available and UAV flight relies on UAV sensors is a potential research topic.